# Electron–Proton Co-doping-Induced Metal–Insulator Transition in VO$_2$ Film via Surface Self-Assembled L-Ascorbic Acid Molecules

*Bowen Li[+], Liyan Xie[+], Zhaowu Wang, Shi Chen, Hui Ren, Yuliang Chen, Chengming Wang, Guobin Zhang, Jun Jiang,\* and Chongwen Zou\**

**Abstract:** *Charge doping is an effective way to induce the metal–insulator transition (MIT) in correlated materials for many important utilizations, which is however practically limited by problem of low stability. An electron–proton co-doping mechanism is used to achieve pronounced phase modulation of monoclinic vanadium dioxide (VO$_2$) at room temperature. Using L-ascorbic acid (AA) solution to treat VO$_2$, the ionized AA$^-$ species donate electrons to the adsorbed VO$_2$ surface. Charges then electrostatically attract surrounding protons to penetrate, and eventually results in stable hydrogen-doped metallic VO$_2$. The variations of electronic structures, especially the electron occupancy of V 3d/O 2p hybrid orbitals, were examined by synchrotron characterizations and first-principle theoretical simulations. The adsorbed molecules protect hydrogen dopants from escaping out of lattice and thereby stabilize the metallic phase for VO$_2$.*

As a typical strongly correlated oxide, vanadium dioxide (VO$_2$) shows a characteristic thermally induced metal–insulator phase transition (MIT) at about 340 K.[1,2] Across the phase-transition boundary, the resistance of VO$_2$ varies up to five orders of magnitude and the infrared transmittance undergoes a pronounced switching effect,[3,4] making VO$_2$ a promising candidate for various applications including memory materials, smart windows, and ultra-fast optical switching devices.[5–7] Many extensive studies have been conducted to modulate VO$_2$ MIT behavior.[8–12] Controlling the charge density is an effective way to modulate competitively electronic phases of VO$_2$. At present, the main control methods include atomic doping[13–15] and the electric-field gating[16–18] with ionic liquid as the dielectric layer. Charge transferring complexes in coordination chemistry can donate charges to VO$_2$ crystal.[19,20] The Iwasa group[21] has induced VO$_2$ MIT under mild conditions by surface adsorption of organic polar molecules. However, most phase modulation effects induced by organic molecules absorptions are weak, as the amounts of injected electrons are limited by the Coulomb screening effect.

The L-ascorbic acid (AA) molecule is known as a good electron donor to transition-metal oxides such as TiO$_2$.[22,23] A recent study[24] reported stabilizing an intermediate metallic state in vanadium dioxide nanobeams via absorption of AA molecules on VO$_2$ wire tips. It was suggested that the AA molecules was chelated on the VO$_2$ top-end and induced charge carrier density reorganization, triggering a metallic state in monoclinic VO$_2$. However, the proposed mechanism was not such convincible if considering the surface charge-transfer-induced Coulomb screening effect. More experimental evidence is desirable for clarifying this charge-doping-induced phase-modulation behavior at room temperature. Furthermore, charge doping in VO$_2$ is normally not stable if no external field is applied. Thus, developing a way to modulate MIT in VO$_2$ and stabilize its metallic phase is a pressing issue.

Herein, we report a facile approach to modulate the phase transition of VO$_2$ film in L-ascorbic acid (AA) or its sodium salt (AA-Na) solutions (Figure 1a and b, details in the Supporting Information). L-AA molecules will be ionized in water as AA$^-$ and H$^+$ ions; The AA$^-$ species bind to the VO$_2$ surface, causing charge transfer owing to the surface coordination effect; electrons flowing into VO$_2$ will electrostatically attract the surrounding free protons in solution to penetrate into the VO$_2$ lattice (Figure 1c); the meet of electrons and protons eventually results in neutral hydrogen atoms inside lattice. Such an electron–proton co-doping strategy thus created a stable H-doping VO$_2$ which exhibits metallic properties at room temperature. Importantly, the adsorbed AA molecules on the surface can protect the hydrogen dopant from evaporation/escaping. Synchrotron-based characterizations, together with first-principle simulations, demonstrated the surface charge transfer and H-doping induced O 2p–V 3d orbitals occupancy variations, which account for the formation of metallic state. These findings would lead to a novel facile way to modulate stable phase transition for correlated oxide materials.

We start with immersing an insulator VO$_2$ film into AA solution at 50 °C or AA-Na solution at room temperature,

[*] B. Li,[+] S. Chen, H. Ren, Y. Chen, G. Zhang, C. Zou
National Synchrotron Radiation Laboratory
University of Science and Technology of China
Hefei, 230029 (China)
E-mail: czou@ustc.edu.cn

L. Xie,[+] C. Wang, Prof. J. Jiang
Hefei National Laboratory for Physical Sciences at the Microscale,
Collaborative Innovation Center of Chemistry for Energy Materials,
CAS Center for Excellence in Nanoscience, School of Chemistry and
Materials Science, University of Science and Technology of China
Hefei, Anhui 230026 (P. R. China)
E-mail: jiangj1@ustc.edu.cn

Z. Wang
School of Physics and Engineering, Henan University of Science and
Technology, Henan Key Laboratory of Photoelectric Energy Storage
Materials and Applications
Luoyang, Henan 471023 (China)

[+] These authors contributed equally to this work.

Supporting information and the ORCID identification number(s) for the author(s) of this article can be found under:
https://doi.org/10.1002/anie.201904148.

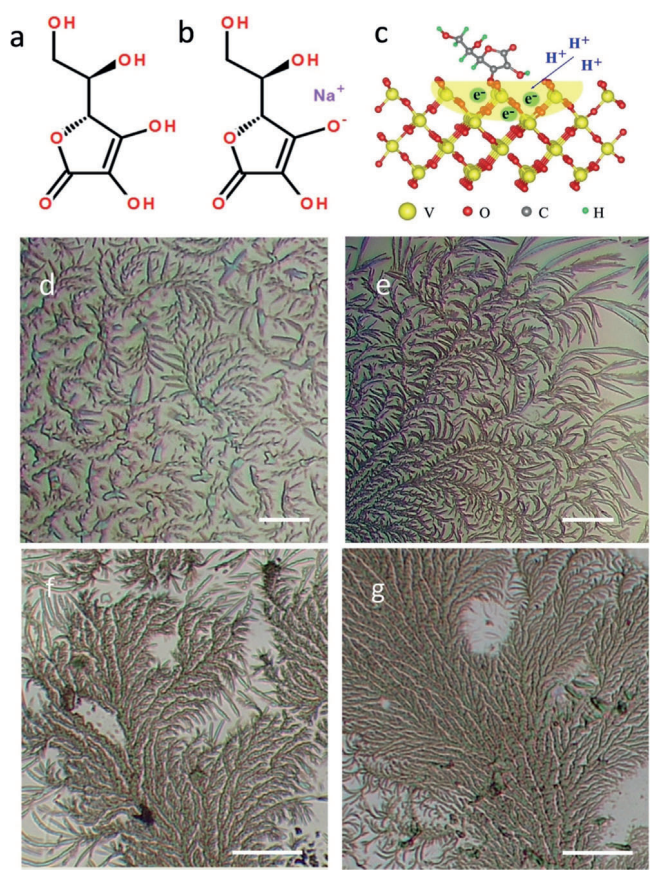

***Figure 1.*** a) L-ascorbic acid (AA) and b) its sodium salt (AA-Na) molecule. c) The surface absorption of AA$^-$ on a VO$_2$ crystal surface. As a strong electron donor, L-AA molecules can connect with the surface V-O to form an organic molecule ligand and electrons are injected from the molecules to VO$_2$ side, which in turn drives surrounding solution protons to penetrate into VO$_2$ due to electrostatic attraction, resulting a stable H atom doping. d)–g) The optical images for the VO$_2$ surface after immersion in AA solution for 5, 30, 60, 120 min, respectively, showing the distinct self-assembled growth of AA molecules on the surface. While it should be noticed that the self-assembled growth is not very uniform since the nucleation is associated with the oxygen vacancies distribution on the surface. Scale bars: d),e) 10 µm, f),g) 50 µm.

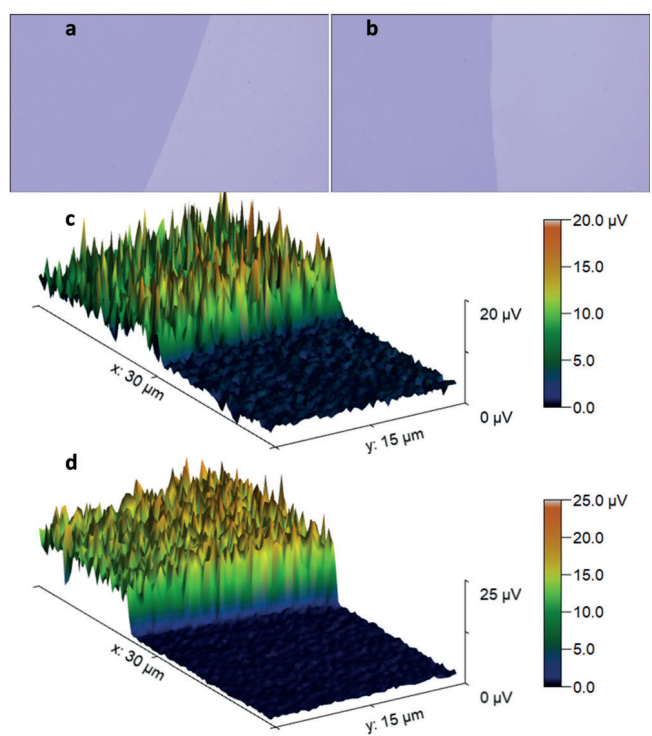

***Figure 2.*** a) Optical images of AA-treated VO$_2$ epitaxial film. Reaction time: 2 h at 50 °C. b) Optical images of AA-Na-treated VO$_2$ epitaxial film and the reaction time is 2 h at 25 °C. c),d) The s-SNOM Images of the VO$_2$ epitaxial film treated with AA or AA-Na at different reaction temperatures for 2 h, with the incident wavelength $\lambda = 10.247$ um.

both of which form a strong acid environment. Figure 1 d–g record the images of VO$_2$ surface after reflux reaction in AA solution at 50 °C with different immersing time. Owing to the crystallization and self-assembled growth of AA molecules on the lattice surface, well-regulated patterns were observed. The surface absorption of AA$^-$ on the VO$_2$ crystal surface act as the initial nucleation site, and consequently the tree-branch like AA nano-crystals are gradually expanded with increasing treatment time until the film surface were densely covered. The surface self-assembled growth behavior is also observed in AA-Na solution at room temperature (Supporting Information, Figure S1).

The optical properties of VO$_2$ with AA/AA-Na were examined by near-field scanning optical microscopy (s-SNOM) with spatial resolution up to 10 nm.[4] In the polarized optical images (Figure 2a and b), the bright domain for the insulating monoclinic phase and the dark domain for the acid treated phase[25] were well-separated. The s-SNOM characterization in Figure 2c and d show that the scattering signals of the dark parts are much stronger than that of the bright parts. Such cliff-like scattering signals in s-SNOM tests suggest the clear boundary for the pristine and treated parts.

Importantly, the VO$_2$ film with self-assembled AA covering can maintain stable metallic phase at room temperature, which is insensitive to temperature varying even after undergoing many resistance–temperature cycling tests (Figure 3a). While if the assembled AA layer is removed by deionized water, the metallic state became very unstable (Figure 3b). It was recovered to the insulating state after several cycles of resistance–temperature tests. The infrared transmittance in photo-absorption spectra (Figure 3c) is significantly reduced by AA treatment, confirming the induced metallic phase. The H atom concentration as a function of VO$_2$ film depth (sputtering time) was investigated by secondary ion mass spectroscopy (SIMS) in Figure 3d, demonstrating the doping of H inside lattice. Furthermore, the AA-Na treated sample shows higher hydrogen concentration than that in AA treated sample, further confirming the better performance of AA-Na treatment owing to its high ionization in solution. This H atom insertion induced the metallic VO$_2$ phase at room temperature (Supporting Information, Figures S2, S3). This AA/AA-Na absorption-induced metallic VO$_2$ film is stable in acid solution, exhibiting excellent anti-corrosion property[26] in 15 wt% H$_2$SO$_4$ solution (Supporting Information, Figure S4) and nearly intact AFM images (Supporting Information, pristine VO$_2$ suggest the coexistence two phases due to the gradual diffusion of H-atoms along the film depth direction.

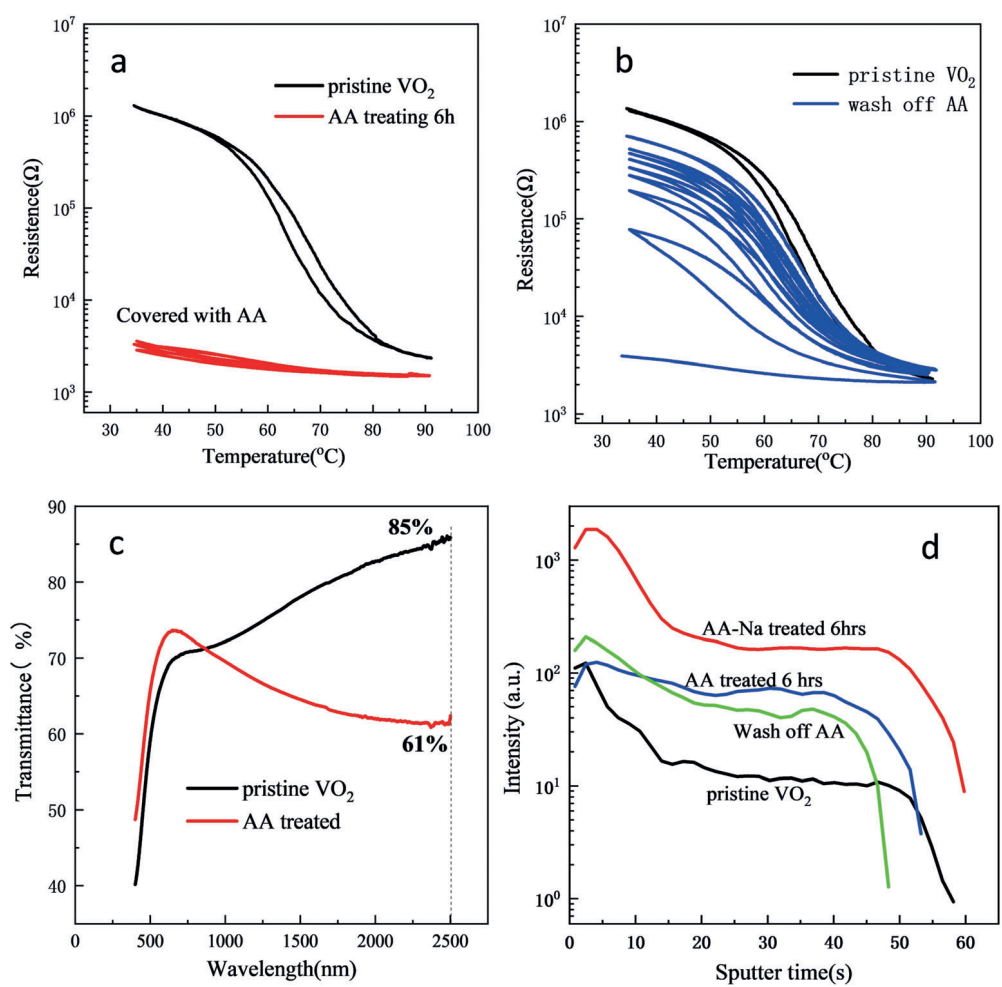

**Figure 3.** a) The electrical measurements for AA treated VO$_2$ samples. The VO$_2$ film covered by assembled AA molecules shows a quite stable metallic state at room temperature. b) If washing away the assembled AA molecules, the metallic VO$_2$ film becomes unstable and it will gradually recovered to the initial insulating monoclinic state after cycled R-T tests. c) The UV/Vis-IR spectra for VO$_2$ film before and after the AA treatment, showing the transitions of metallic states at room temperature. d) The SIMS spectra for H atom concentration as the function of film depth (sputtering time) for the pristine VO$_2$ film and AA or AA-Na treated samples.

The chemical states and electronic structures were characterized by XPS and XANES in Figure 4c and d. The XPS peaks were calibrated by the reference line of C 1s peak at 284.8 eV. For the pristine VO$_2$ sample, the V 2p$_{3/2}$ feature consisted a dominant V$^{4+}$ peak and a smaller V$^{5+}$ peak according to the curve fitting by V$^{4+}$ peak at 516.2 eV and V$^{5+}$ peak at 517.5 eV.[27,28] The smaller V$^{5+}$ peak is originated from surface oxidation. While after L-AA-treatment, the V 2p$_{3/2}$ peak became much broader, which can be fitted by three peaks due to the new appearance of V$^{3+}$ related feature at 515.0 eV.[29] Comparison of the V 2p$_{3/2}$ peaks before and after AA treatment indicated that though most vanadium in the thin film still showed +4 valence state, some +3 valence state was produced due to H-doping. XANES spectra in Figure 4d showed the variation of V L-edge and O K-edge absorption, in which the distinct peak shift for V-L$_{III}$ and V-L$_{II}$ edges towards low energy direction confirmed the reduced valance state of V atoms due to AA treatment. The t$_{2g}$/e$_g$ absorption peak of O K-edge in Figure 4d give the information of electron occupancy on the V 3d and O 2p hybrid orbitals. The O K-edge was related to the transition between O 1s and O 2p level. The intensity ratio of t$_{2g}$/e$_g$ decreased with the AA treatment, reflecting that t$_{2g}$ levels of d$_\parallel$* and π* orbitals were gradually occupied by electrons coming from H-dopants.[30]

First-principles theoretical calculations at the density functional theory (DFT) level were performed using the Vienna ab initio simulation package (VASP).[31] The adsorption of L-ascorbic acid molecules on VO$_2$ (010) surface may take four configurations (Supporting Information, Figure S7). The adsorption energies of AA and ionized AA$^-$ on the oxygen vacancy (O$_V$) defect surface of VO$_2$ (AA/VO$_2$(O$_V$) and AA$^-$/VO$_2$(O$_V$)) are −0.96 eV and −2.41 eV, respectively. While those of AA and ionized AA$^-$ on the perfect surface of

Figure S5). Furthermore, the H-doped VO$_2$ films exhibit no sharp MIT behavior even within the low temperature range (Supporting Information, Figure S6).

XRD and Raman spectra were measured to examine H-doping induced structural variations (Figure 4). The strong Al$_2$O$_3$ (0001) substrate diffraction peak at 41.68° and VO$_2$ (020) peak at 39.8° (JCPDS# 82-0661) were observed. The VO$_2$ (020) peak show pronounced shifts towards low-angle directions after AA treatment, indicating the expanding of the crystal lattice along the (020) direction owing to hydrogen intercalation. The sample treated by 6 h shows the diffraction peak at about 39.6° which is readily indexed to tetragonal rutile VO$_2$(R) structure (JCPDS# 79-1655). Raman spectra in Figure 4b shows the typical characteristic peaks at 191 cm$^{-1}$, 221 cm$^{-1}$, 308 cm$^{-1}$, 618 cm$^{-1}$ for the pristine insulating states, all of which disappear after 6 h treatment (Figure 4b), further confirming the crystal structure transformation to tetragonal rutile-like structure (metallic state).[18,26,27] With less treating time of about 2 h, several smaller characteristic peaks for $VO_2$ (AA/$VO_2$ and $AA^-$/$VO_2$) are $-1.43$ eV and $-1.96$ eV, respectively. This means that ionized $AA^-$ adsorb on the $VO_2$ ($O_V$) surface is the most stable case (Supporting Information, Table S1). The computed differential charge distribution at the molecule–$VO_2$ interfaces in those four adsorption configurations suggested the transferring of 0.10–0.57 electrons per unit from molecules to $VO_2$ (Figure 5 a–d; Supporting Information, Table S1). These are further confirmed by the calculated molecular-adsorbate induced interfacial states occupancies of 0.24, 0.46, 0.24, and 0.57 electrons per unit for four different configurations (Supporting Information, Figure S8). These negative electrons will drive protons ($H^+$) to penetrate into $VO_2$ (Figure 1 c), resulting in neutral H-dopants.[26] The migration path of H from surface to subsurface were revealed (Supporting Information, Figure S9). Electrons in $VO_2$ indeed lower down the migration barrier from 3.40 eV in neutral $VO_2$ to 3.05 eV in $VO_2$ with one electron (Figure 5 e; Supporting Information, Table S2). The formation energy of an oxygen vacancy defect on the $VO_2$ surface is increased from 4.16 eV in neutral $VO_2$ to 4.47 eV in that with an electron (Supporting Information, Table S2), suggesting better structural stability. The formed O−H covalent bond with a length of 0.99 Å causes subtle distortions to the lattice structure (Supporting Information, Figure S10), while charge analysis found polarization charges of 0.67 $e^-$ being donated from the H-dopant to the lattice unit. These polarization charges change the occupation of electronic states, as reflected in the density of state (DOS) (Figure 5 f), explaining the metallic property found in H-doped $VO_2$ samples.

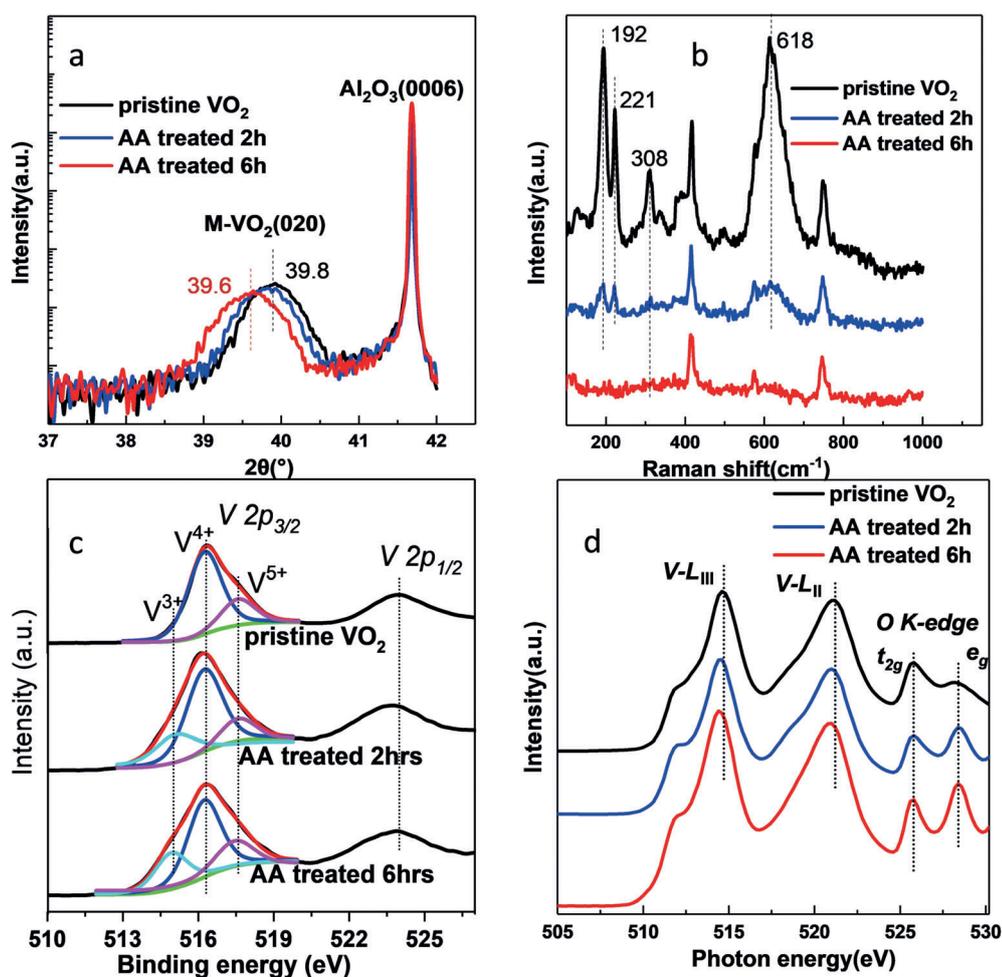

**Figure 4.** a) The XRD patterns for the AA treated $VO_2$ films with different treating time, clear diffraction peak shifts towards low angle directions can be observed. b) Raman spectra for the AA treated $VO_2$ films. With the increasing treating time, the clear Raman peaks from monoclinic $VO_2$ film gradually disappear. c) The XPS spectra for $VO_2$ films before and after the AA treatment. d) The XANES spectra for AA treated $VO_2$ films.

In summary, we have achieved stabilized metallic phase of monoclinic $VO_2$ at room temperature by the surface absorption of AA molecules. The surface coordination-effect-induced charge transferring from $AA^-$ species into $VO_2$ side, while electrons attract protons in solution to penetrate into $VO_2$ lattice. These constitute a novel electron–proton co-doping mechanism, which forms the effective hydrogen doping and induces MIT for $VO_2$. The self-assembled growth of AA molecules can strength the hydrogen doping and maintain stable H-doped metallic $VO_2$ at room temperature. This new mechanism is certified by various synchrotron-based characterizations as well as the DFT calculations. It is believed that the surface coordination effect may provide a new perspective to regulate electronic properties of many correlated materials or two-dimensional electronic materials.

### Acknowledgements

This work was partially supported by the National Key Research and Development Program of China (2016YFA0401004, 2018YFA0208603), the National Natural Science Foundation of China (11404095, 11574279, 11704362), the funding supported by the Youth Innovation Promotion Association CAS, the Major/Innovative Program of Development Foundation of Hefei Center for Physical Science and Technology and the China Postdoctoral Science Foundation (2017M622002). This work was partially carried out at the USTC Center for Micro and Nanoscale Research and Fabrication. The authors also acknowledge supports from the XMCD beamline (BL12B) in National Synchrotron Radiation Laboratory (NSRL) of Hefei.

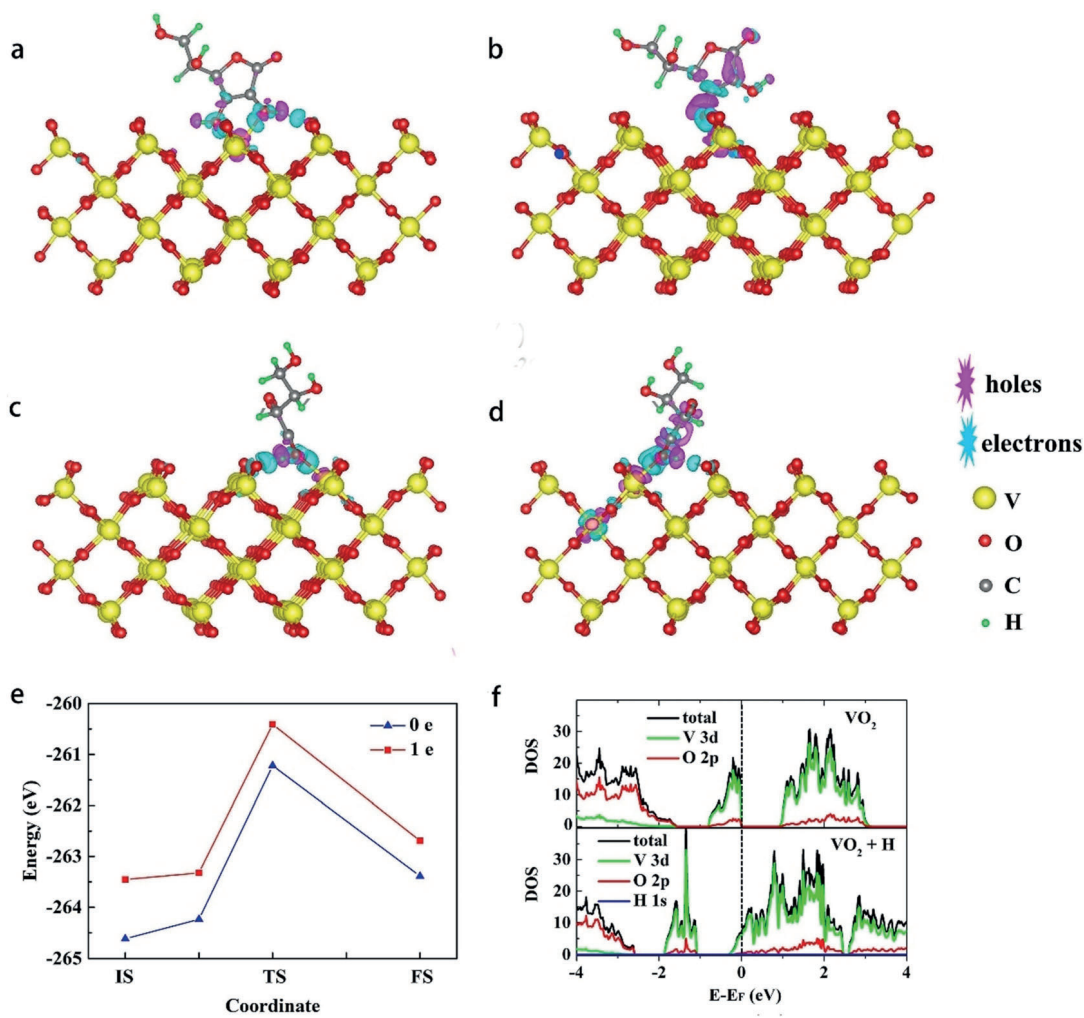

*Figure 5.* a) Computed differential charge distribution at the interfaces of AA molecule on defective $VO_2$ surface ($AA/VO_2(O_V)$). b) Ionized $AA^-$ molecule defective $VO_2$ surface ($AA^-/VO_2(O_V)$). c) AA molecule on perfect $VO_2$ surface ($AA/VO_2$). d) ionized $AA^-$ molecule and perfect $VO_2$ surface($AA^-/VO_2$). e) The energy variations due to H migration from the $VO_2$ surface to sub-surface. f) Computed density of states (DOS) of pristine and H-doping monoclinic $VO_2$.

## Conflict of interest

The authors declare no conflict of interest.

**Keywords:** first-principle simulations · hydrogenation · metal–insulator phase transition · synchrotron characterization

[1] F. J. Morin, *Phys. Rev. Lett.* **1959**, *3*, 34–36.
[2] T. Yao, L. Liu, C. Xiao, X. Zhang, Q. Liu, S. Wei, Y. Xie, *Angew. Chem. Int. Ed.* **2013**, *52*, 7554–7558; *Angew. Chem.* **2013**, *125*, 7702–7706.
[3] Z. Li, Y. Guo, Z. Hu, J. Su, J. Zhao, J. Wu, J. Wu, Y. Zhao, C. Wu, Y. Xie, *Angew. Chem. Int. Ed.* **2016**, *55*, 8018–8022; *Angew. Chem.* **2016**, *128*, 8150–8154.
[4] M. M. Qazilbash, M. Brehm, B.-G. Chae, P. C. Ho, G. O. Andreev, B.-J. Kim, S. J. Yun, A. V. Balatsky, M. B. Maple, F. Keilmann, H.-T. Kim, D. N. Basov, *Science* **2007**, *318*, 1750.
[5] T. Driscoll, H.-T. Kim, B.-G. Chae, B.-J. Kim, Y.-W. Lee, N. M. Jokerst, S. Palit, D. R. Smith, M. Di Ventra, D. N. Basov, *Science* **2009**, *325*, 1518.
[6] X. Li, R. E. Schaak, *Angew. Chem. Int. Ed.* **2017**, *56*, 15550–15554; *Angew. Chem.* **2017**, *129*, 15756–15760.
[7] C. Wu, F. Feng, Y. Xie, *Chem. Soc. Rev.* **2013**, *42*, 5157–5183.
[8] R. M. Wentzcovitch, W. W. Schulz, P. B. Allen, *Phys. Rev. Lett.* **1994**, *72*, 3389–3392.
[9] T. Yao, X. Zhang, Z. Sun, S. Liu, Y. Huang, Y. Xie, C. Wu, X. Yuan, W. Zhang, Z. Wu, G. Pan, F. Hu, L. Wu, Q. Liu, S. Wei, *Phys. Rev. Lett.* **2010**, *105*, 226405.
[10] J. H. Park, J. M. Coy, T. S. Kasirga, C. Huang, Z. Fei, S. Hunter, D. H. Cobden, *Nature* **2013**, *500*, 431.
[11] J. D. Budai, J. Hong, M. E. Manley, E. D. Specht, C. W. Li, J. Z. Tischler, D. L. Abernathy, A. H. Said, B. M. Leu, L. A. Boatner, R. J. McQueeney, O. Delaire, *Nature* **2014**, *515*, 535.

[12] V. R. Morrison, R. P. Chatelain, K. L. Tiwari, A. Hendaoui, A. Bruhács, M. Chaker, B. J. Siwick, *Science* **2014**, *346*, 445.
[13] C. Piccirillo, R. Binions, I. P. Parkin, *Thin Solid Films* **2008**, *516*, 1992–1997.
[14] X. Lv, Y. Cao, L. Yan, Y. Li, Y. Zhang, L. Song, *ACS Appl. Mater. Interfaces* **2018**, *10*, 6601–6607.
[15] B. Fisher, *J. Phys. Chem. Solids* **1982**, *43*, 205–211.
[16] J. Jeong, N. Aetukuri, T. Graf, T. D. Schladt, M. G. Samant, S. S. P. Parkin, *Science* **2013**, *339*, 1402.
[17] Y. Hou, R. Xiao, X. Tong, S. Dhuey, D. Yu, *Nano Lett.* **2017**, *17*, 7702–7709.
[18] J. Wei, H. Ji, W. Guo, A. H. Nevidomskyy, D. Natelson, *Nat. Nanotechnol.* **2012**, *7*, 357.
[19] H. Zhou, J. Li, Y. Xin, X. Cao, S. Bao, P. Jin, *J. Mater. Chem. C* **2015**, *3*, 5089–5097.
[20] K. Wang, W. Zhang, L. Liu, P. Guo, Y. Yao, C.-H. Wang, C. Zou, Y.-W. Yang, G. Zhang, F. Xu, *Appl. Surf. Sci.* **2018**, *447*, 347–354.
[21] H. Shioya, Y. Shoji, N. Seiki, M. Nakano, T. Fukushima, Y. Iwasa, *Appl. Phys. Express* **2015**, *8*, 121101.
[22] T. Rajh, J. M. Nedeljkovic, L. X. Chen, O. Poluektov, M. C. Thurnauer, *J. Phys. Chem. B* **1999**, *103*, 3515–3519.
[23] Y. Ou, J.-D. Lin, H.-M. Zou, D.-W. Liao, *J. Mol. Catal. A* **2005**, *241*, 59–64.
[24] Z. Li, J. Wu, Z. Hu, Y. Lin, Q. Chen, Y. Guo, Y. Liu, Y. Zhao, J. Peng, W. Chu, C. Wu, Y. Xie, *Nat. Commun.* **2017**, *8*, 15561.
[25] J. Cao, E. Ertekin, V. Srinivasan, W. Fan, S. Huang, H. Zheng, J. W. L. Yim, D. R. Khanal, D. F. Ogletree, J. C. Grossman, J. Wu, *Nat. Nanotechnol.* **2009**, *4*, 732.
[26] Y. Chen, Z. Wang, S. Chen, H. Ren, L. Wang, G. Zhang, Y. Lu, J. Jiang, C. Zou, Y. Luo, *Nat. Commun.* **2018**, *9*, 818.
[27] J. Lin, H. Ji, M. W. Swift, W. J. Hardy, Z. Peng, X. Fan, A. H. Nevidomskyy, J. M. Tour, D. Natelson, *Nano Lett.* **2014**, *14*, 5445–5451.
[28] M. Demeter, M. Neumann, W. Reichelt, *Surf. Sci.* **2000**, *454–456*, 41–44.
[29] G. Silversmit, D. Depla, H. Poelman, G. B. Marin, R. De Gryse, *J. Electron Spectrosc. Relat. Phenom.* **2004**, *135*, 167–175.
[30] J. Mendialdua, R. Casanova, Y. Barbaux, *J. Electron Spectrosc. Relat. Phenom.* **1995**, *71*, 249–261.
[31] S. Chen, Z. Wang, L. Fan, Y. Chen, H. Ren, H. Ji, D. Natelson, Y. Huang, J. Jiang, C. Zou, *Phys. Rev. B* **2017**, *96*, 125130.

## Communications

**Phase Transitions**

B. Li, L. Xie, Z. Wang, S. Chen, H. Ren, Y. Chen, C. Wang, G. Zhang, J. Jiang,* C. Zou* ■■■■–■■■■

Electron–Proton Co-doping-Induced Metal–Insulator Transition in $VO_2$ Film via Surface Self-Assembled L-Ascorbic Acid Molecules

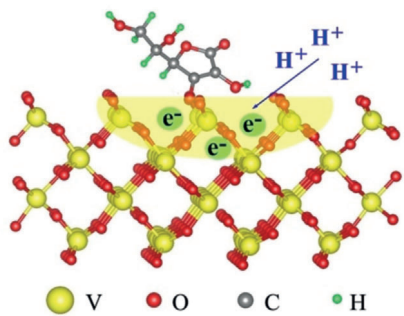

**An electron–proton** co-doping mechanism is used to achieve pronounced phase modulation of monoclinic $VO_2$ at room temperature. Using an L-ascorbic acid (AA) solution to treat $VO_2$, the ionized $AA^-$ species donate electrons to the adsorbed $VO_2$ surface. Charges then electrostatically attract surrounding protons to penetrate, and eventually results in stable hydrogen-doped metallic $VO_2$.